\documentclass[aps,prd,twocolumn,showpacs,preprintnumbers,floatfix]{revtex4-2}

\usepackage{amsmath,amssymb,amsfonts}
\usepackage{graphicx}
\usepackage{booktabs}
\usepackage{hyperref}

\newcommand{\diff}{\mathrm{d}}
\newcommand{\ee}{\mathrm{e}}
\newcommand{\ii}{\mathrm{i}}
\newcommand{\Tr}{\mathrm{Tr}}
\newcommand{\Nefold}{\mathcal{N}}
\newcommand{\Rnoise}{\mathcal{R}}

\begin{document}

\title{Inflationary Branch Decoherence and the Cosmological Arrow of Time}

\author{Ali Nayeri}
\email{ali.nayeri@ordinalresearch.org}
\affiliation{Clear Quantum Corporation, Lewes, Delaware 19958, USA}
\affiliation{Ordinal Research Institute, Wilmington, DE 19801, USA}

\date{\today}

\begin{abstract}
We analyze branch decoherence in inflationary quantum cosmology by
computing reduced density matrices and branch-overlap factors for
long-wavelength perturbations.  The Hartle--Hawking no-boundary state
is real in the semiclassical regime and contains both expanding and
contracting WKB components, whereas the tunneling state is selected as
an outgoing complex WKB branch; expanding--contracting decoherence is
therefore central for the former and mainly diagnostic for the latter.
Using the influence-functional formalism, we derive the noise kernel for
a light spectator environment and evaluate decoherence under
horizon-based and EFT-motivated coarse grainings.  We then compute the
single-mode branch overlap directly from the Bunch--Davies mode
functions, obtaining $|\mathcal{D}_k(z)|=[z^2/(z^2+1)]^{1/4}$ in the
massless limit and $|\mathcal{D}_k(z)|\sim z^\nu$ on superhorizon scales
for massive fields, where $z\equiv-k\eta$ is the dimensionless
wavenumber with $\eta$ the conformal time.  In the massless case the
accumulated geometric
branch functional is evaluated in closed form, with a leading
cutoff-sensitive phase-space term and a universal subleading
contribution.  The calculation provides an explicit quantitative bridge
between quantum-cosmological boundary conditions, inflationary
squeezing, and the emergence of effectively classical cosmological
histories.
\end{abstract}


\maketitle

\section{Quantum cosmology before classicality}
\label{sec:intro}

\subsection{Decoherence and the classical limit in quantum cosmology}

Quantum cosmology seeks to explain how a universal quantum state gives
rise to the effectively classical spacetime described by cosmological
observations.  The Wheeler--DeWitt equation and its associated boundary
conditions define a wavefunctional of geometry and matter, but they do
not by themselves explain why observations are well described by
classical backgrounds with stochastic perturbations.

This issue is sharpened during inflation.  Semiclassical WKB structure
can identify approximate background trajectories, and squeezing can
make correlation functions appear classical, yet neither effect by
itself suppresses phase coherence between macroscopically distinct
alternatives.  The quantum-to-classical transition therefore requires a
dynamical mechanism acting on the reduced state of the observable
degrees of freedom.

The natural framework for that problem is coarse graining and
environment-induced decoherence.  By tracing over inaccessible degrees
of freedom one obtains an influence functional for the long-wavelength
variables, with dissipation and noise kernels that govern the loss of
phase coherence and the onset of effectively classical stochastic
dynamics.  In this sense, classical spacetime is not put in by hand but
emerges from a reduced description of a larger quantum state.

The aim of this paper is to make the dynamical origin of classical
spacetime explicit by deriving the decoherence functional for
inflationary curvature perturbations from the underlying action.  Our
focus is not only on the general claim that decoherence occurs, but on
three more specific questions: how the effective long-mode coupling to a
spectator sector arises, how the decoherence rate depends on the coarse
graining, and how branch decoherence clarifies the relation between
quantum-cosmological boundary conditions and classical spacetime
histories.

More concretely, the universal state is described by a wavefunctional
\begin{equation}
\Psi[h_{ij}(\mathbf{x}),\Phi(\mathbf{x})],
\label{eq:universal_wavefunctional}
\end{equation}
defined over three-geometries and matter field configurations.  In
practice one works in a truncated configuration space consisting of
minisuperspace variables, such as the scale factor~$a$, together with
quantum perturbations.  Standard no-boundary and tunneling boundary
conditions~\cite{HartleHawking,Vilenkin} then lead, in the semiclassical
regime, to WKB-type branch expansions of the form
\begin{equation}
\Psi[a,\zeta,\sigma] \simeq
  \sum_{\alpha} A_{\alpha}(a)\,\ee^{\ii S_{\alpha}(a)}\,
  \psi_{\alpha}[\zeta,\sigma;a],
\end{equation}
where $\alpha$ labels distinct semiclassical histories.  This structure
supports approximate classical trajectories and approximate
Schr\"{o}dinger evolution for perturbations along each branch, but it
does not by itself suppress interference between macroscopically
distinct alternatives.  The physically relevant object is therefore the
reduced density matrix obtained by tracing over unobserved variables,
defined explicitly in equation~(\ref{eq:rho_reduced}); classical
spacetime emerges only if this reduced state becomes approximately
diagonal in an appropriate basis~\cite{Zeh,JoosZeh,Kiefer,
HalliwellHawking1985,GellMannHartle1990,Schlosshauer2007}.

Recent work has further emphasized the role of environment-induced
decoherence in inflationary cosmology.  In particular, Burgess, Holman,
and Hoover showed that interactions between long-wavelength perturbations
and subhorizon environmental degrees of freedom can generate efficient
decoherence of inflationary fluctuations, even when the underlying
dynamics remains unitary~\cite{BurgessHolman}.  Their analysis
demonstrated that the decoherence rate is controlled by the coupling
between system and environment as well as the rapid growth of physical
phase space during inflation.  Our treatment is complementary to these
approaches: rather than focusing primarily on phenomenological estimates
within an inflationary EFT, we derive the effective spectator-field
coupling from the underlying action, construct the associated
influence-functional noise kernel, and use the resulting branch-overlap
structure to separate the role of boundary-condition amplitudes from the
dynamical emergence of classical histories.

The relation between decoherence and the arrow of time in quantum
cosmology also has a substantial history.  Kiefer's early treatment of
decoherence in quantum cosmology already discussed the loss of
coherence between expanding and contracting WKB components of the
universal wavefunction~\cite{Kiefer1992}, while Zeh's monograph gives a
standard account of how time-asymmetric classical records can emerge
from special cosmological boundary conditions and environmental
decoherence~\cite{ZehBook}.  Recent reviews and conference proceedings
summarize the broader literature on time, semiclassical branches, and
decoherence in quantum cosmology~\cite{Chataignier2024}.  The purpose
of the present paper is therefore not to claim that branch decoherence
or the arrow-of-time problem is new.  The new element is the explicit
inflationary branch-overlap calculation, including its closed
single-mode form and the analytic evaluation of the accumulated
geometric branch functional.

\medskip\noindent\textit{What is new here.}---
While decoherence of inflationary perturbations has been discussed in
earlier work (e.g.\ \cite{BurgessHolman,KieferPolarski}), the present
analysis makes five specific contributions.
(i)~We derive the effective long-wavelength spectator coupling directly
from the underlying covariant action in comoving gauge (rather than
postulating it within an EFT) and identify the corresponding
influence-functional noise-kernel structure.
(ii)~We compare horizon-based and effective-field-theory coarse
grainings within a common framework and show that both lead to robust
growth of the decoherence functional.
(iii)~We evaluate the branch-overlap factor $|\mathcal{D}_k(z)|$
explicitly using the Bunch--Davies mode functions, obtaining the exact
closed form $|\mathcal{D}_k(z)|=[z^2/(z^2+1)]^{1/4}$ in the massless
limit and the power law $|\mathcal{D}_k(z)|\propto z^\nu$ for massive
fields, where $z\equiv-k\eta$ denotes the dimensionless wavenumber
($\eta$ the conformal time, with $z\gg1$ subhorizon and $z\to0$
superhorizon).  In the massless limit this exact single-mode result leads to a
closed-form evaluation of the accumulated branch functional
$\Gamma_{+-}$ and an analytic asymptotic decomposition into a
cutoff-sensitive leading term and a universal subleading piece.  To our
knowledge this explicit evaluation has not appeared previously: earlier
works characterised classicality through squeezing parameters (Polarski
and Starobinsky~\cite{PolarskiStarobinsky}), reduced-density-matrix
elements (Kiefer and Polarski~\cite{KieferPolarski}), or noise-kernel
rates (Burgess, Holman and Hoover~\cite{BurgessHolman}), but did not
extract the closed-form branch-overlap factor or the exact massless
accumulated functional.
(iv)~We derive the dissipation kernel explicitly and show that it is
consistent with the de~Sitter fluctuation-dissipation relation at
Gibbons--Hawking temperature $T_{\mathrm{GH}}=H/(2\pi)$; the resulting
Langevin equation is consistent with Starobinsky's stochastic inflation
in the superhorizon limit, once the noise normalization is matched to
the standard Bunch--Davies result.
(v)~We explicitly separate boundary-condition amplitudes from dynamical
branch-overlap suppression, thereby clarifying the different roles of
the Hartle--Hawking and Vilenkin proposals versus environment-induced
decoherence in the emergence of an operational cosmological arrow.  As a further
corollary, we show that the same exact single-mode overlap law applies
to tensor mode functions in exact de~Sitter, establishing a limited but
useful overlap-law universality without entering a separate tensor
decoherence-rate analysis.

\begin{figure*}[t]
\centering
\includegraphics[width=0.85\textwidth,trim={0pt 0pt 0pt 6.5pt},clip]{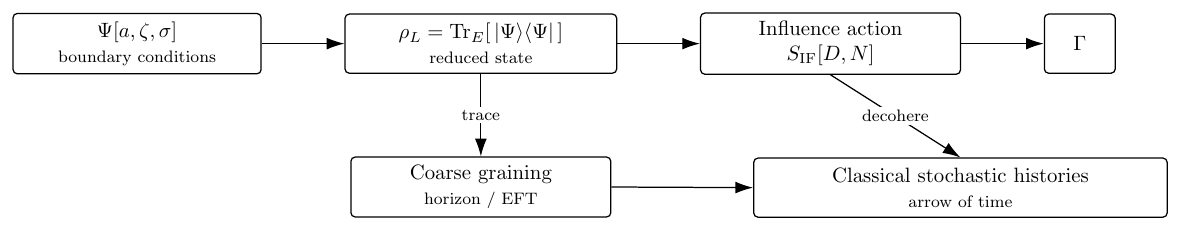}
\caption{Schematic roadmap of the emergence of classical cosmological
histories from quantum-cosmological boundary conditions via
influence-functional
decoherence~\cite{FeynmanVernon,CaldeiraLeggett}.
Boundary conditions define the universal state~$\Psi$.  Tracing over
unobserved degrees of freedom yields a reduced density matrix for
long-wavelength perturbations.  The influence functional encodes
dissipation and noise; the associated decoherence functional suppresses
interference between macroscopically distinct histories.  Classical
stochastic descriptions and an operational arrow of time emerge after
coarse graining, with the detailed rate depending only weakly on the
coarse-graining prescription.}
\label{fig:roadmap}
\end{figure*}

\section{Reduced density matrix and influence functional}
\label{sec:influence}

We separate the degrees of freedom into long-wavelength curvature
perturbations~$\zeta_L$ (the system) and unobserved degrees of freedom,
including short-wavelength modes~$\zeta_S$ and additional
fields~$\sigma$ (the environment).  The reduced density matrix is
\begin{equation}
\rho_L[a,\zeta_L;\,a',\zeta_L']
  = \int \mathcal{D}\zeta_S\,\mathcal{D}\sigma\;
    \Psi[a,\zeta_L,\zeta_S,\sigma]\,
    \Psi^{*}[a',\zeta_L',\zeta_S,\sigma].
\label{eq:rho_reduced}
\end{equation}

The time evolution of~$\rho_L$ can be written using the
Schwinger--Keldysh
formalism~\cite{FeynmanVernon,CaldeiraLeggett,CalzettaHu1994,HuMataczPazSinha1993}.
Integrating out the environment defines an influence functional whose
general structure is encoded in the influence action
$\mathcal{F}=\ee^{\ii S_{\mathrm{IF}}}$, where
\begin{equation}
\begin{aligned}
S_{\mathrm{IF}}
&= \int \diff^{4}x\,\diff^{4}x'\,
\Bigl[
  \zeta_{\Delta}(x)\,D(x,x')\,\zeta_{c}(x')
  \\
&\quad
  + \frac{\ii}{2}\,
    \zeta_{\Delta}(x)\,N(x,x')\,\zeta_{\Delta}(x')
\Bigr] + \cdots .
\end{aligned}
\label{eq:influence_action}
\end{equation}
Here $D$ is a real, causal dissipation kernel and $N$ is a real,
positive-semidefinite noise kernel.  Decoherence is governed by the
imaginary term involving~$N$.

The suppression of off-diagonal elements of the reduced density matrix
is controlled by the decoherence functional
\begin{equation}
\Gamma[\zeta_{\Delta}]
  = \frac{1}{2}
    \int \diff^{4}x\,\diff^{4}x'\;
    \zeta_{\Delta}(x)\,N(x,x')\,\zeta_{\Delta}(x').
\label{eq:decoherence_functional}
\end{equation}
When $\Gamma \gg 1$ for macroscopically distinct configurations, the
reduced density matrix becomes approximately diagonal.

\textit{A note on two distinct decoherence functionals.}
This paper computes $\Gamma$ in two complementary ways that yield
different e-fold scales and should not be confused.

\emph{(i)~Geometric branch-decoherence functional}
(section~\ref{sec:boundary}).  The \emph{single-mode} overlap factor
$|\mathcal{D}_k(z)|$ is independent of both the spectator
coupling~$\hat\lambda$ and the UV cutoff: it is determined entirely by
the Bunch--Davies mode functions on the two semiclassical branches.
The \emph{band-integrated} diagnostic
$\Gamma_{+-}=-\sum_k\ln|\mathcal{D}_k|$ necessarily depends on the
upper phase-space boundary~$k_{\max}$ once one sums over a finite mode
band, but this dependence is analytically controlled: in the massless
limit the integral can be evaluated in exact closed form
(section~\ref{sec:boundary}), and the resulting asymptotic expansion
separates $\Gamma_{+-}$ into a cutoff-sensitive leading term and a
universal subleading piece.  The integrated functional crosses unity
within $\approx 0.5$ $e$-folds for representative mode bands.

\emph{(ii)~Interaction-dependent noise-kernel functional}
(sections~\ref{sec:kernel} and~\ref{subsec:coarsegraining}).  This
measures the rate at which a specific spectator sector suppresses
off-diagonal elements of the reduced density matrix for the system
perturbation~$\zeta_L$.  It depends explicitly on the spectator
coupling~$\hat\lambda$ and, through the noise kernel~$N(x,x')$, on the
UV regulation scheme; this functional crosses unity at several
to~$\sim\!10$ $e$-folds depending on~$\hat\lambda$ and on the coarse
graining.

The two quantities are not competing predictions for the same process:
the first isolates the geometric asymmetry between cosmological
branches --- with a cutoff-independent single-mode law and an
analytically controlled band-integrated diagnostic --- while the second
quantifies how system--environment interactions exploit that asymmetry,
with rates that depend on the noise-kernel regulation.  Both confirm
irreversible classicalisation well within inflation.

\section{Explicit noise kernel for a light environment field}
\label{sec:kernel}

We consider an environment consisting of a light spectator scalar
field~$\sigma$ with mass $m_{\sigma}\lesssim H$ in the Bunch--Davies
(BD) vacuum~\cite{PolarskiStarobinsky,Kiefer,Maldacena2003,CheungEtAl2008}.
Rather than postulating the system--environment coupling, we derive its
leading long-wavelength form from the covariant spectator action
\begin{equation}
S_{\sigma} = -\frac{1}{2}
  \int \diff^4x\,\sqrt{-g}
  \left[g^{\mu\nu}\partial_\mu\sigma\,\partial_\nu\sigma
        + m_\sigma^2\,\sigma^2\right].
\label{eq:sigma_action_covariant}
\end{equation}
Working in comoving gauge, where the inflaton fluctuation is set to
zero and the scalar metric perturbation is carried by~$\zeta$,
\begin{align}
h_{ij}&=a^2 e^{2\zeta}\delta_{ij},\nonumber\\
N &= 1+\frac{\dot\zeta}{H}+\mathcal{O}(\zeta^2,\partial^2),\qquad
N_i = \partial_i\psi,
\label{eq:adm_comoving_gauge}
\end{align}
the spectator action becomes, using the ADM decomposition,
\begin{equation}
\begin{aligned}
S_{\sigma}
&= \frac{1}{2}\int \diff t\,\diff^3x\,N\sqrt{h}\,
  \Bigl[
    N^{-2}\!\left(\dot{\sigma}-N^i\partial_i\sigma\right)^{\!2}
    \\
&\hspace{3.0em}
    - h^{ij}\partial_i\sigma\,\partial_j\sigma
    - m_\sigma^2\,\sigma^2
\Bigr].
\end{aligned}
\label{eq:sigma_action_adm}
\end{equation}
Expanding to first order in the long-wavelength curvature perturbation
gives
\begin{equation}
S_{\sigma} = S_{\sigma}^{(0)} + S_{\sigma\zeta}^{(3)} + \cdots,
\label{eq:sigma_split}
\end{equation}
with
\begin{equation}
\begin{aligned}
S_{\sigma\zeta}^{(3)}
&= \frac{1}{2}\int \diff t\,\diff^3x\,a^3
   \Biggl[
     \left(3\zeta+\frac{\dot\zeta}{H}\right)\dot\sigma^2
     \\
&\quad
     -\left(\zeta+\frac{\dot\zeta}{H}\right)
      \frac{(\partial_i\sigma)^2}{a^2}
     \\
&\quad
     -\left(3\zeta+\frac{\dot\zeta}{H}\right)
      m_\sigma^2\,\sigma^2
   \Biggr] + \cdots .
\end{aligned}
\label{eq:zeta_sigma_cubic_full}
\end{equation}
where the ellipsis denotes terms proportional to $N_i$ and
slow-roll-suppressed constraint corrections.
As discussed in Maldacena's detailed treatment of the comoving-gauge
constraints~\cite{Maldacena2003}, the shift~$N_i$ is determined by the
momentum constraint and its contribution at this order enters only
through constraint-suppressed terms, which is why the $N_i$ sector can
be consistently grouped with the slow-roll-suppressed corrections in
equation~(\ref{eq:zeta_sigma_cubic_full}).
Equation~(\ref{eq:zeta_sigma_cubic_full}) makes clear that the coupling
is not an arbitrary ansatz: a long-wavelength~$\zeta_L$ locally
rescales the spectator Hamiltonian density.  For the superhorizon system
mode of interest, $\dot\zeta_L/H=\mathcal{O}(\epsilon)\zeta_L$ is
slow-roll suppressed, and if the spectator is light and dominated by its
mass term on sufficiently long wavelengths,
equation~(\ref{eq:zeta_sigma_cubic_full}) reduces to the local form
\begin{equation}
S_{\mathrm{int}}^{\mathrm{eff}}
= -\frac{3}{2}\,m_\sigma^2
  \int \diff t\,\diff^3x\,a^3\,\zeta_L\,\sigma^2
\equiv
  -\int \diff t\,\diff^3x\,a^3\,\lambda_{\mathrm{eff}}\,\zeta_L\,\sigma^2,
\label{eq:Sint_zeta_sigma2}
\end{equation}
with $\lambda_{\mathrm{eff}}=\tfrac{3}{2}m_\sigma^2$ in the minimal
spectator model and with additional $\mathcal{O}(\epsilon H^2)$ or
derivative corrections in more general EFTs of inflation.  In conformal
time, $\diff t\,a^3=\diff\eta\,a^4$, reproducing the scaling assumed in
the influence-functional treatment.  This reduction should be viewed as
the appropriate effective coupling in the mass-dominated long-wavelength
regime.  For environmental modes that remain deep inside the horizon,
$q/a\gg H\gtrsim m_\sigma$, the gradient and kinetic terms retained in
equation~(\ref{eq:zeta_sigma_cubic_full}) dominate over the mass term,
so the subhorizon contribution to decoherence is more generally
described by the full operator structure of
equation~(\ref{eq:zeta_sigma_cubic_full}) rather than solely by the
reduced coupling~(\ref{eq:Sint_zeta_sigma2}).  The coarse-grained
estimates below should therefore be interpreted as parametric estimates
of decoherence growth, with equation~(\ref{eq:Sint_zeta_sigma2})
supplying the cleanest explicit coupling in the long-wavelength sector.

For a Gaussian environment state, the noise kernel appearing in
equation~(\ref{eq:decoherence_functional})
is~\cite{FeynmanVernon,CaldeiraLeggett}
\begin{equation}
N(x,x') =
  \lambda_{\mathrm{eff}}^{2}\,
  \mathrm{Re}\!\left[G^{>}_{\sigma}(x,x')^{2}\right],
\label{eq:noise_kernel_G2}
\end{equation}
where $G^{>}_{\sigma}(x,x')=\langle\sigma(x)\sigma(x')\rangle$ is the
Wightman function.  Using BD mode functions one finds
\begin{equation}
G^{>}_{\sigma}(x,x')
  = \int\frac{\diff^{3}k}{(2\pi)^{3}}\,
    \ee^{\ii\mathbf{k}\cdot(\mathbf{x}-\mathbf{x}')}\,
    u_{k}(\eta)\,u^{*}_{k}(\eta'),
\label{eq:wightman_def}
\end{equation}
with
\begin{equation}
u_{k}(\eta)
  = \frac{\sqrt{\pi}}{2}\,
    \ee^{\ii(\nu+1/2)\pi/2}\,H(-\eta)^{3/2}\,
    H^{(1)}_{\nu}(-k\eta),
\label{eq:bd_modefunction}
\end{equation}
with
\begin{equation}
\nu = \sqrt{\frac{9}{4}-\frac{m_{\sigma}^{2}}{H^{2}}},
\label{eq:nu_def}
\end{equation}
where $H^{(1)}_{\nu}$ is a Hankel function.  For a light field it is
convenient to define
\begin{equation}
\Delta = \frac{3}{2}-\nu \simeq \frac{m_{\sigma}^{2}}{3H^{2}},
\label{eq:Delta_def}
\end{equation}
which controls the superhorizon time dependence.

Substituting
equations~(\ref{eq:wightman_def})--(\ref{eq:bd_modefunction}) into
equation~(\ref{eq:noise_kernel_G2}) yields the explicit convolution
\begin{equation}
\begin{aligned}
N(\eta,\eta';\mathbf{k})
&= \lambda_{\mathrm{eff}}^{2}\,
   \mathrm{Re}\!\Biggl[
      \int\frac{\diff^{3}q}{(2\pi)^{3}}\,
      u_{q}(\eta)\,u^{*}_{q}(\eta')
      \\
&\hspace{4.1em}\times
      u_{|\mathbf{k}-\mathbf{q}|}(\eta)\,
      u^{*}_{|\mathbf{k}-\mathbf{q}|}(\eta')
    \Biggr].
\end{aligned}
\label{eq:noise_kernel_convolution}
\end{equation}
For a superhorizon long mode $k_L\ll aH$, the integral is dominated by
$q\gg k_L$ and simplifies to
\begin{equation}
\begin{aligned}
N(\eta,\eta';k_L)
&\simeq \lambda_{\mathrm{eff}}^{2}
    \int\frac{\diff^{3}q}{(2\pi)^{3}}
    \\
&\quad\times
    \mathrm{Re}\!\bigl[
      \bigl(u_{q}(\eta)\,u^{*}_{q}(\eta')\bigr)^{2}
    \bigr].
\end{aligned}
\label{eq:noise_kernel_superhorizon}
\end{equation}

In the superhorizon regime $-q\eta\ll1$, the mode functions scale as
$u_q(\eta)\propto(-\eta)^{\Delta}\,q^{-3/2+\Delta}$, implying
\begin{equation}
N(\eta,\eta';k_L)
  \propto \lambda_{\mathrm{eff}}^{2}\,H^{4}\,
          (-\eta)^{2\Delta}(-\eta')^{2\Delta}
          \int\diff q\,q^{-4+4\Delta}.
\label{eq:noise_kernel_scaling}
\end{equation}
This estimate refers to the infrared contribution to the kernel; the
horizon-based and EFT-based decoherence estimates below are instead
dominated by environmental modes that remain subhorizon over the
relevant coarse-graining range.  A nonzero mass $m_{\sigma}\neq0$
softens the infrared behavior, although in the light-field regime the
integral still requires a physical infrared cutoff associated with the
finite inflationary patch.  Writing
\begin{equation}
\int_{q_{\min}}\diff q\,q^{-4+4\Delta}
  \propto \frac{q_{\min}^{-3+4\Delta}}{-3+4\Delta},
\quad \Delta\neq\tfrac{3}{4},
\label{eq:ir_cutoff_regulated}
\end{equation}
shows explicitly that for any finite $q_{\min}$ the kernel is finite
and acquires a regulated dependence on $m_\sigma/H$ through~$\Delta$.
The derivation also shows that the coefficient controlling decoherence
is fixed by the spectator-sector Hamiltonian density and therefore has
a definite mass dimension; in the minimal model
$\lambda_{\mathrm{eff}}\propto m_\sigma^2$, while in a more general EFT
it receives higher-derivative and slow-roll-suppressed corrections.

\medskip\noindent\textit{Special case $\Delta=3/4$.}---
Equation~(\ref{eq:ir_cutoff_regulated}) excludes the value
$\Delta=3/4$, corresponding to $\nu=3/4$ and
$m_\sigma = \tfrac{3\sqrt{3}}{4}H\approx 1.30\,H$.  At this mass the
integrand $q^{-4+4\Delta}=q^{-1}$ is logarithmically divergent and
equation~(\ref{eq:ir_cutoff_regulated}) is replaced by
\begin{equation}
\int_{q_{\min}}^{q_{\max}}\diff q\,q^{-1}
  = \ln\!\left(\frac{q_{\max}}{q_{\min}}\right),
\quad \Delta = \tfrac{3}{4}.
\label{eq:ir_log_case}
\end{equation}
The accumulated branch-decoherence functional therefore grows
logarithmically in the size of the inflationary patch rather than as a
power law, while the overlap factor itself still vanishes superhorizon
as $|\mathcal{D}_k|\propto z^{3/4}$ --- verified numerically to better
than $0.03\%$ by comparing $|\mathcal{D}_k(z_1)|/|\mathcal{D}_k(z_2)|$
with $(z_1/z_2)^{3/4}$ at $z_1=0.003$, $z_2=0.010$, using the general
formula~(\ref{eq:D_overlap_result}) with Hankel functions evaluated at
$\nu=3/4$.  This case lies at the boundary between power-law and
logarithmic infrared behavior and is the analogue of the
Breitenlohner--Freedman saturating mass for a light field in de~Sitter.

\subsection{Dissipation kernel and the stochastic inflation limit}
\label{subsec:dissipation}

The influence action~(\ref{eq:influence_action}) contains both a noise
kernel $N$ and a dissipation kernel $D$.  While the noise kernel governs
decoherence and has been the focus of the preceding analysis, $D$
determines the back-reaction of the environment on the system dynamics
and must be computed to obtain the full open-system effective action.
We derive $D$ here and show that the resulting Langevin equation for
$\zeta_L$ is consistent with Starobinsky's stochastic
inflation~\cite{StarobinskyStochastic,NambuSasaki1989}
in the superhorizon limit, once the noise normalization is matched to
the standard result.

For a Gaussian environment state, the dissipation kernel follows from
the imaginary part of the same two-point correlator that gives $N$:
\begin{equation}
D(x,x') = \lambda_{\mathrm{eff}}^{2}\,\theta(\eta-\eta')\,
  \mathrm{Im}\!\left[G^{>}_\sigma(x,x')^{2}\right].
\label{eq:dissipation_kernel_def}
\end{equation}
The $\theta(\eta-\eta')$ factor enforces causality: $D$ is a retarded
kernel, in contrast to the symmetric noise kernel $N$.

In the WKB approximation for subhorizon environmental modes
($q\gg aH$), the Bunch--Davies mode functions reduce to
$u_q(\eta)\simeq(2q)^{-1/2}e^{-iq\eta}$, so
$(u_q u_q^*)^2=(4q^2)^{-1}e^{-2iq\delta\eta}$ with $\delta\eta\equiv\eta-\eta'$.
Substituting into equations~(\ref{eq:noise_kernel_G2})
and~(\ref{eq:dissipation_kernel_def}) and integrating over the
environment ($q>k_c$) with an implicit UV convergence factor
$e^{-\varepsilon q}$ ($\varepsilon\to0^+$) to regulate the oscillatory tail,
one obtains the two kernels in momentum space
\begin{align}
N(\eta,\eta';k_L)
  &= \frac{\lambda_{\mathrm{eff}}^{2}}{64\pi^{2}}
     \frac{\sin\!\bigl[2k_c(\eta-\eta')\bigr]}{\eta-\eta'},
\label{eq:N_explicit_wkb}\\[6pt]
D(\eta,\eta';k_L)
  &= \frac{\lambda_{\mathrm{eff}}^{2}}{64\pi^{2}}\,
     \theta(\eta-\eta')\,
     \frac{\cos\!\bigl[2k_c(\eta-\eta')\bigr]}{\eta-\eta'},
\label{eq:D_explicit_wkb}
\end{align}
where $k_c = \epsilon\,a(\eta)H$ is the coarse-graining cutoff.  The
two kernels differ by a $90^\circ$ phase shift: $N$ involves
$\sin[2k_c\delta\eta]$ while $D$ involves $\cos[2k_c\delta\eta]$, with
$\delta\eta\equiv\eta-\eta'$.  These are the standard Caldeira--Leggett
kernels for an ohmic bath with a sharp lower cutoff~\cite{CaldeiraLeggett,CalzettaHu1994}.

\medskip\noindent\textit{Fluctuation-dissipation relation.}---
In the long-wavelength limit $|\omega|\ll k_c$, the Fourier transforms
of equations~(\ref{eq:N_explicit_wkb})--(\ref{eq:D_explicit_wkb}) are
consistent with the de~Sitter fluctuation-dissipation relation
\begin{align}
\tilde{D}(\omega)
  &= -\frac{\ii\omega}{2T_{\mathrm{GH}}}\,\tilde{N}(\omega),
\nonumber\\
T_{\mathrm{GH}} &= \frac{H}{2\pi},
\label{eq:fdr_desitter}
\end{align}
where $T_{\mathrm{GH}}$ is the Gibbons--Hawking temperature of
de~Sitter space~\cite{GibbonsHawking1977}.  This KMS-type relation is
a general property of quantum fields in de~Sitter and holds for the
exact Bunch--Davies two-point function; the WKB kernels above are
consistent with the interpretation that the open-system dynamics is
tied to the horizon thermodynamics at $T_{\mathrm{GH}}=H/(2\pi)$.

\medskip\noindent\textit{Stochastic equation of motion.}---
The influence action with both $N$ and $D$ generates a Langevin
equation for the long-wavelength curvature perturbation,
\begin{equation}
\mathcal{P}[\zeta_L](\eta)
  + \Gamma\,\partial_\eta\zeta_L(\eta)
  = \xi(\eta),
\label{eq:langevin_zeta}
\end{equation}
where $\mathcal{P}$ is the free equation-of-motion operator for
$\zeta_L$ in de~Sitter (which reduces to $-\epsilon H^2\zeta_L$ at
leading order in slow roll on superhorizon scales), $\xi(\eta)$ is a
Gaussian stochastic noise satisfying
\begin{align}
\langle\xi(\eta)\rangle &= 0,
\nonumber\\
\langle\xi(\eta)\,\xi(\eta')\rangle &= N(\eta,\eta';k_L),
\label{eq:noise_correlator}
\end{align}
and the friction coefficient is estimated in the Markovian
approximation.  The integral $\int_0^\infty D(\delta\eta)\diff\delta\eta$
is oscillatory and requires UV regularization; with a physical
UV cutoff $q_{\max}\gg k_c$, the leading contribution is
\begin{equation}
\Gamma
  \simeq \frac{\lambda_{\mathrm{eff}}^{2}}{128\pi^{2}k_c}
  = \frac{\hat\lambda^{2}H}{128\pi^{2}\epsilon},
\label{eq:friction_coefficient}
\end{equation}
where the $k_c$ in the denominator arises because the dominant
contribution to the Markovian integral comes from the scale
$\delta\eta\sim1/(2k_c)$ set by the coarse-graining cutoff.
The numerical prefactor in equation~(\ref{eq:friction_coefficient})
is regulator-dependent at order unity; the robust content is the
scaling $\Gamma\propto\lambda_{\mathrm{eff}}^2/k_c$.
Since $\Gamma/H = \hat\lambda^{2}/(128\pi^{2}\epsilon)\ll 1$ for
$\hat\lambda\ll 1$ and $\epsilon\sim\mathcal{O}(1)$, the friction term
is negligible on superhorizon scales and $\zeta_L$ is effectively
frozen.  In the superhorizon limit $\mathcal{P}[\zeta_L]\to0$.  Matching the
noise normalization to the standard Bunch--Davies vacuum fluctuation
amplitude, i.e.\ replacing the WKB-derived kernel with the local
approximation
$N(\eta,\eta')\approx(H^2/4\pi^2)\,a^{-4}\,\delta(\eta-\eta')$,
the Langevin equation~(\ref{eq:langevin_zeta}) reduces to
\begin{equation}
\diff\zeta_L \simeq \frac{H}{2\pi}\,\diff W(\Nefold),
\label{eq:stochastic_inflation}
\end{equation}
where $\diff W$ is a Wiener increment per e-fold.  This is
Starobinsky's stochastic inflation equation~\cite{StarobinskyStochastic,NambuSasaki1989}.
The influence-functional framework therefore provides a consistent
open-system setting in which the decoherence calculation above and the
classical stochastic description of inflationary perturbations coexist,
although the noise normalization in the Langevin equation is fixed by
matching to the standard Bunch--Davies result rather than derived
directly from the WKB kernels~(\ref{eq:N_explicit_wkb})--(\ref{eq:D_explicit_wkb}).

\subsection{Coarse graining and decoherence rates}
\label{subsec:coarsegraining}

In this section we evaluate the decoherence functional derived above
under two physically motivated coarse-graining prescriptions, following
the general logic developed in the cosmological decoherence
literature~\cite{PolarskiStarobinsky,BurgessHolman,KieferPolarski,
MartinVenninPeter2012,BurgessHolmanStamou2022}.
While the precise numerical rate of decoherence depends on the coarse
graining, its qualitative behavior---namely the efficient suppression
of interference between macroscopically distinct curvature
perturbations---is robust.

\subsection{Horizon-based coarse graining}

A natural prescription in inflationary cosmology is to define the
environment as modes whose physical wavelength is shorter than the
Hubble radius at a given time.  Concretely, we split modes according to
\begin{equation}
q > k_{c}(\eta) \equiv \epsilon\,a(\eta)\,H,
\end{equation}
where $\epsilon\ll1$ is a fixed numerical parameter.  Modes
continuously cross from the environment into the system as inflation
proceeds, but decoherence is generated primarily before horizon
crossing, when environmental modes are still oscillatory.

For subhorizon modes $q\gg aH$, the BD mode functions admit a WKB
approximation,
\begin{equation}
u_{q}(\eta) \simeq \frac{1}{\sqrt{2q}}\,\exp(-\ii q\eta),
\end{equation}
so that
\begin{equation}
\left(u_{q}(\eta)\,u_{q}^{*}(\eta')\right)^{2}
  \simeq \frac{1}{4q^{2}}\,\exp[-2\ii q(\eta-\eta')].
\end{equation}
Substituting into the noise kernel yields
\begin{equation}
N(\eta,\eta';k_L)
  \simeq \frac{\lambda_{\mathrm{eff}}^{2}}{4}
         \int_{q>k_{c}}\frac{\diff^{3}q}{(2\pi)^{3}}
         \frac{\exp[-2\ii q(\eta-\eta')]}{q^{2}}.
\end{equation}
The rapid oscillations render the kernel sharply peaked near
$\eta=\eta'$, allowing the local approximation
$N(\eta,\eta')\simeq\Rnoise(\eta)\,\delta(\eta-\eta')$.
The decoherence functional for a frozen superhorizon mode then becomes
\begin{equation}
\Gamma_{k_L}
  \simeq \frac{1}{2}\,|\Delta\zeta_{k_L}|^{2}
         \int\diff\eta\,a^{4}(\eta)\,\Rnoise(\eta).
\end{equation}
Changing variables to the number of e-folds $\Nefold=\ln a$ gives
the parametric estimate
\begin{equation}
\frac{\diff\Gamma_{k_L}}{\diff\Nefold}
  \sim \frac{\lambda_{\mathrm{eff}}^{2}}{H^{4}}\,
       |\Delta\zeta_{k_L}|^{2}\,a^{3}(\Nefold),
\end{equation}
up to numerical factors of order unity.  Decoherence therefore
accumulates monotonically, and once $\Gamma_{k_L}\gg1$ the reduced
density matrix becomes effectively diagonal.

\subsection{EFT-based coarse graining}

An alternative prescription is motivated by effective field theory.
One introduces a fixed physical cutoff $\Lambda_{\mathrm{phys}}$
satisfying $H\ll\Lambda_{\mathrm{phys}}\ll M_{\mathrm{Pl}}$, and
defines the environment as modes with physical momenta above this
scale, $q>a(\eta)\,\Lambda_{\mathrm{phys}}$.  Modes traced out in this
way never re-enter the system, rendering decoherence effectively
irreversible.

For all such modes $q/a\gg H$, and the WKB approximation remains valid
throughout inflation.  The noise kernel again becomes local in time,
$N(\eta,\eta') \simeq \Rnoise_{\Lambda}\,\delta(\eta-\eta')$,
with
\begin{equation}
\Rnoise_{\Lambda}
  \sim \lambda_{\mathrm{eff}}^{2}
       \int_{q>a\Lambda_{\mathrm{phys}}}
       \frac{\diff^{3}q}{(2\pi)^{3}}\frac{1}{q^{2}}
  \sim \lambda_{\mathrm{eff}}^{2}\,\Lambda_{\mathrm{phys}}\,a(\eta),
\end{equation}
giving
\begin{equation}
\frac{\diff\Gamma_{k_L}}{\diff\Nefold}
  \sim \frac{\lambda_{\mathrm{eff}}^{2}\,\Lambda_{\mathrm{phys}}}{H^{5}}\,
       |\Delta\zeta_{k_L}|^{2}\,a^{4}(\Nefold).
\end{equation}
Decoherence is thus even more efficient than in the horizon-based
prescription and remains insensitive to the details of ultraviolet
completion.  In both schemes, interactions with unobserved degrees of
freedom rapidly suppress interference between distinct long-wavelength
curvature perturbations, providing a robust dynamical origin for
classicality.

\subsection{Order-of-magnitude estimates}
\label{subsec:estimates}

Observations of the cosmic microwave background imply $A_s\simeq
2.1\times10^{-9}$, giving
$\zeta_{\mathrm{rms}}\sim\sqrt{A_s}\sim5\times10^{-5}$.
Macroscopically distinct semiclassical branches may be characterized
by $|\Delta\zeta_{k_L}|\sim n\,\zeta_{\mathrm{rms}}$ with
$n\sim10$--$100$.

Using the EFT-based coarse graining,
\begin{equation}
\Gamma_{k_L}(\Delta\Nefold)
  \sim \frac{\hat\lambda^2}{4}\,
       \frac{\Lambda_{\mathrm{phys}}}{H}\,
       |\Delta\zeta_{k_L}|^2\,
       \bigl(\ee^{4\Delta\Nefold}-1\bigr),
\end{equation}
where $\hat\lambda\equiv\lambda_{\mathrm{eff}}/H^2$ and $\Delta\Nefold$
counts e-folds since horizon exit.  Setting
$\Gamma_{k_L}(\Delta\Nefold_{\mathrm{dec}})\simeq1$ gives
\begin{equation}
\Delta\Nefold_{\mathrm{dec}}
  = \frac{1}{4}
    \ln\!\left[
      1+\frac{4}{\hat\lambda^2\,(\Lambda_{\mathrm{phys}}/H)\,
                 |\Delta\zeta_{k_L}|^2}
    \right].
\label{eq:decoherence_efolds_exact}
\end{equation}

For $\Lambda_{\mathrm{phys}}/H\sim10$, the explicit values are given in
table~\ref{tab:decoherence_times}.  In the minimal spectator model,
$\hat\lambda=\tfrac{3}{2}m_\sigma^2/H^2$, so, for example,
$m_\sigma/H=0.1$ gives $\hat\lambda=0.015$ and
$\Delta\Nefold_{\mathrm{dec}}\approx5.67$ ($n=10$) and $4.52$
($n=100$); $m_\sigma/H=0.3$ gives $\hat\lambda=0.135$ and $4.57$
($n=10$) and $3.42$ ($n=100$).  Decoherence occurs within a few to at
most $\mathcal{O}(10)$ e-folds across the full range, confirming
efficient classicalisation well before the end of inflation.

\begin{table}[t]
\caption{Decoherence time $\Delta\Nefold_{\mathrm{dec}}$ (e-folds after
horizon exit) from equation~(\ref{eq:decoherence_efolds_exact}), for
$\Lambda_{\mathrm{phys}}/H=10$ and branch separation
$|\Delta\zeta_{k_L}|=n\,\zeta_{\mathrm{rms}}$ with
$\zeta_{\mathrm{rms}}\simeq5\times10^{-5}$.}
\label{tab:decoherence_times}
\begin{ruledtabular}\begin{tabular}{@{}lcc@{}}
$\hat\lambda$
  & $\Delta\Nefold_{\mathrm{dec}}$ $(n=10)$
  & $\Delta\Nefold_{\mathrm{dec}}$ $(n=100)$ \\
\hline
$10^{-5}$ & 9.33 & 8.18 \\
$10^{-3}$ & 7.03 & 5.88 \\
$10^{-1}$ & 4.72 & 3.57 \\
\end{tabular}\end{ruledtabular}
\end{table}

\section{Hartle--Hawking and tunneling proposals after decoherence}
\label{sec:boundary}

We now return to the interpretation of quantum-cosmological boundary
conditions in light of the decoherence mechanism developed above.
Both the no-boundary proposal of Hartle and Hawking and the tunneling
proposal of Vilenkin define pure quantum states of the universe and,
by themselves, do not yield classical spacetimes.  Decoherence provides
the missing dynamical ingredient.
There is, however, an important distinction between the two boundary
conditions.  The Hartle--Hawking state is real and, in the
semiclassical WKB regime, has the standing-wave form associated with a
superposition of expanding and contracting components.  The Vilenkin
tunneling state is instead selected as a complex outgoing WKB wave,
schematically $\Psi_{\mathrm{T}}\propto\exp(\ii S)$, and already
chooses the expanding branch in the usual semiclassical interpretation.
Consequently, decoherence between expanding and contracting branches is
physically required to obtain an ensemble of classical alternatives for
the Hartle--Hawking state, whereas for the tunneling state the same
calculation is best viewed as a diagnostic of the stability of the
selected outgoing branch against its time-reversed counterpart.

\subsection{Semiclassical branch structure}

In the semiclassical regime, the universal wavefunction admits a WKB
decomposition
\begin{equation}
\Psi[a,\zeta,\sigma] \simeq
  \sum_{\alpha} A_{\alpha}(a)\,\exp\!\left(\ii S_{\alpha}(a)\right)
  \psi_{\alpha}[\zeta,\sigma;a],
\end{equation}
where $\alpha$ distinguishes different semiclassical branches.  For
homogeneous minisuperspace variables these branches typically correspond
to expanding and contracting classical solutions, while for
perturbations they encode different inflationary histories.  The
Wheeler--DeWitt equation is time-reversal invariant, so the
semiclassical equation itself admits both orientations.  The
Hartle--Hawking boundary condition keeps them in a real WKB
superposition, while the tunneling condition imposes an outgoing complex
combination~\cite{HartleHawking,Vilenkin}.  In the no-boundary case the
branches coexist coherently at the level of the full wavefunction; the
WKB approximation alone does not turn this superposition into a
classical ensemble.

\subsection{Reduced density matrix in the branch basis}

Projecting the reduced density matrix onto the semiclassical branch
basis yields matrix elements of the schematic form
\begin{equation}
\rho_{\alpha\beta}
  \equiv \int\mathcal{D}\zeta_S\,\mathcal{D}\sigma\,
         \Psi_{\alpha}\,\Psi^{*}_{\beta}
  \propto A_{\alpha}\,A^{*}_{\beta}\,
         \exp\!\left[-\Gamma_{\alpha\beta}\right],
\end{equation}
where $\Gamma_{\alpha\beta}$ is the decoherence functional for the
difference between the long-wavelength field configurations associated
with branches $\alpha$ and~$\beta$.  When $\Gamma_{\alpha\beta}\gg1$,
the off-diagonal elements are exponentially suppressed and the density
matrix becomes approximately diagonal.  Only at this stage does it
become meaningful to interpret the wavefunction as an ensemble of
mutually exclusive classical histories.

\subsection{Expanding versus contracting histories}

The reduced density matrix element between the two WKB branches can be
written as
\begin{equation}
\rho_{+-}(a)
  = A_+(a)\,A_-^*(a)\,
    \ee^{\ii M(S_+(a)-S_-(a))}\,\mathcal{D}(a),
\label{eq:branch_overlap_general}
\end{equation}
where
\begin{equation}
\mathcal{D}(a)\equiv
  \Tr_{\mathrm{env}}\!\left[
    U_+(a)\,\rho_{\mathrm{env}}\,U_-^\dagger(a)
  \right],
\label{eq:branch_overlap_D}
\end{equation}
where $U_\pm$ denote the environmental evolution operators conditioned
on the expanding and contracting WKB backgrounds, and $M$ is the large
semiclassical parameter multiplying the gravitational Hamilton--Jacobi
action.  We use $\mathcal{D}$ for the branch-overlap factor throughout
this section to distinguish it from the dissipation kernel~$D$
in~(\ref{eq:influence_action}).  For Gaussian environmental states the
decoherence factor factorises over modes,
\begin{equation}
\mathcal{D}(a)
  = \prod_{k\in\mathrm{env}}\mathcal{D}_k(a)
  = \exp\!\left[-\Gamma_{+-}(a)+\ii\Theta_{+-}(a)\right],
\label{eq:branch_decoherence_factor}
\end{equation}
with $\Gamma_{+-}(a)=-\sum_{k\in\mathrm{env}}\ln|\mathcal{D}_k(a)|$.
Once $\Gamma_{+-}\gg1$, interference between the two semiclassical
geometries is exponentially suppressed.

For an environmental oscillator mode whose conditional states on the
two branches are Gaussian with kernels $\Omega_k^{(+)}$ and
$\Omega_k^{(-)}$,
\begin{equation}
|\mathcal{D}_k|
  = \left[
      \frac{4\,\mathrm{Re}\,\Omega_k^{(+)}\,\mathrm{Re}\,\Omega_k^{(-)}}
           {|\Omega_k^{(+)}+\Omega_k^{(-)\,*}|^2}
    \right]^{1/4},
\label{eq:gaussian_overlap_mode}
\end{equation}
so branch decoherence is controlled by how differently the two
backgrounds squeeze the same environmental mode.  We now evaluate
$|\mathcal{D}_k|$ explicitly using the Bunch--Davies mode functions
already established in section~\ref{sec:kernel}.

\subsubsection*{Gaussian kernels on expanding and contracting
backgrounds}

The vacuum state on each branch is Gaussian:
$\Psi^\pm_k[\phi_k]\propto\exp(-\frac{1}{2}\Omega_k^{(\pm)}|\phi_k|^2)$,
with the kernel related to the mode function by
\begin{equation}
\Omega_k^{(\pm)} = \ii\,\frac{\partial_\eta u_k^{(\pm)}}{u_k^{(\pm)}}.
\label{eq:Omega_from_u}
\end{equation}
On the \emph{expanding} branch ($a=-1/(H\eta)$, $\eta<0$), the
Bunch--Davies solution is
$u_k^{(+)}\propto(-\eta)^{3/2}H^{(1)}_\nu(-k\eta)$.  On the
\emph{contracting} branch ($a=+1/(H\eta)$, $\eta>0$), the
corresponding vacuum is
$u_k^{(-)}\propto\eta^{3/2}H^{(2)}_\nu(k\eta)$.  Using the Hankel
recurrence $(H^{(1)}_\nu)'(z)=H^{(1)}_{\nu-1}(z)-(\nu/z)H^{(1)}_\nu(z)$
with $z=-k\eta$, equation~(\ref{eq:Omega_from_u}) gives
\begin{align}
\Omega_k^{(+)}
  &= -\ii k\!\left[\frac{\Delta}{z}
    + \frac{H^{(1)}_{\nu-1}(z)}{H^{(1)}_\nu(z)}\right],
\label{eq:Omega_plus}\\
\Omega_k^{(-)}
  &= +\ii k\!\left[\frac{\Delta}{z'}
    + \frac{H^{(2)}_{\nu-1}(z')}{H^{(2)}_\nu(z')}\right],
\label{eq:Omega_both}
\end{align}
where $z'=k\eta$ and $\Delta=3/2-\nu$.  The sign reversal reflects the
opposite orientation of conformal time on the two branches.
Throughout, $z=-k\eta>0$ is the dimensionless wavenumber on the
expanding branch ($\eta<0$), while $z'=k\eta>0$ is its counterpart on
the contracting branch ($\eta>0$); they are numerically equal at equal
$|k\eta|$, so the comparison in
equations~(\ref{eq:ReOmega})--(\ref{eq:ImOmega}) is made at equal $|z|$.

\subsubsection*{Real and imaginary parts}

Writing $H^{(1)}_\nu=J_\nu+\ii Y_\nu$, defining
$\mathcal{W}(z)\equiv J_\nu^2(z)+Y_\nu^2(z)>0$, and using the
standard cross-order Wronskian (Abramowitz \& Stegun 9.1.16)
$J_\nu Y_{\nu-1} - J_{\nu-1}Y_\nu = 2/(\pi z)$, one finds
\begin{align}
\mathrm{Re}\,\Omega_k^{(+)}
  &= +\frac{2k}{\pi z\,\mathcal{W}(z)},
\label{eq:ReOmega}\\[4pt]
\mathrm{Im}\,\Omega_k^{(+)}
  &= -k\!\left[\frac{\Delta}{z}
    + \frac{J_{\nu-1}J_\nu+Y_{\nu-1}Y_\nu}{\mathcal{W}(z)}\right].
\label{eq:ImOmega}
\end{align}
Note that $\mathrm{Re}\,\Omega_k^{(+)}>0$ for all $z$, as required for
the Gaussian wavefunction $\Psi^+_k\propto\exp(-\tfrac{1}{2}\Omega_k^{(+)}|\phi_k|^2)$
to be normalizable.
Since $H^{(2)}_\nu=(H^{(1)}_\nu)^*$, the contracting branch satisfies
$\mathrm{Re}\,\Omega_k^{(-)}=\mathrm{Re}\,\Omega_k^{(+)}$ and
$\mathrm{Im}\,\Omega_k^{(-)}=-\mathrm{Im}\,\Omega_k^{(+)}$ at equal
$|k\eta|$.

\subsubsection*{Closed-form result}

Setting $R\equiv\mathrm{Re}\,\Omega_k^{(+)}$ and
$I\equiv\mathrm{Im}\,\Omega_k^{(+)}$, and noting
$\Omega_k^{(+)}+\Omega_k^{(-)\,*}=2R+2\ii I$, substitution into
equation~(\ref{eq:gaussian_overlap_mode}) gives
\begin{equation}
\boxed{
|\mathcal{D}_k(z)|
  = \left[\frac{R^2}{R^2+I^2}\right]^{1/4}
  = \frac{1}{(1+\tan^2\!\theta_k)^{1/4}},
}
\label{eq:D_overlap_result}
\end{equation}
where $\theta_k\equiv\arg\Omega_k^{(+)}$ is the squeezing angle of the
expanding-branch vacuum.  The overlap depends only on the
expanding-branch state; the contracting branch enters through the
symmetry $\mathrm{Im}\,\Omega^{(-)}=-\mathrm{Im}\,\Omega^{(+)}$.

\medskip\noindent\textit{Massless case ($\nu=3/2$, exact).}---
For $\nu=3/2$ the half-integer Bessel functions give
$\mathcal{W}=(2/\pi z)(1+z^{-2})$.  Using
$H^{(1)}_{1/2}(z)=-i\sqrt{2/(\pi z)}e^{iz}$ and
$H^{(1)}_{3/2}(z)=-\sqrt{2/(\pi z)}e^{iz}(1+i/z)$ one finds
$H^{(1)}_{1/2}/H^{(1)}_{3/2}=iz/(z+i)$, so
equation~(\ref{eq:Omega_plus}) gives
$\Omega_k^{(+)}=-ik\cdot iz/(z+i)=kz(z-i)/(z^2+1)$, yielding
\begin{equation}
R = +\frac{kz^2}{z^2+1} > 0, \quad I = -\frac{kz}{z^2+1},
\end{equation}
and the exact closed form
\begin{equation}
|\mathcal{D}_k(z)|\big|_{\nu=3/2}
  = \left[\frac{z^2}{z^2+1}\right]^{1/4}.
\label{eq:D_massless_exact}
\end{equation}
The intermediate step is immediate:
$R^2/(R^2+I^2)=(k^2z^4/(z^2+1)^2)/[(k^2z^4+k^2z^2)/(z^2+1)^2]
=z^2/(z^2+1)$, confirming the closed form.
This satisfies $|\mathcal{D}_k|\to1$ as $z\to\infty$ (subhorizon) and
$|\mathcal{D}_k|\sim z^{1/2}\to0$ as $z\to0$ (superhorizon).

\medskip\noindent\textit{Massive fields ($\nu<3/2$).}---
Superhorizon asymptotics give $\mathcal{W}\sim Y_\nu^2\propto z^{-2\nu}$,
so $R\propto z^{2\nu}\to0$ while $I$ remains finite, giving
\begin{equation}
|\mathcal{D}_k(z)| \sim z^\nu \to 0
\quad (z\to 0),
\label{eq:D_massive_powerlaw}
\end{equation}
verified numerically to better than $0.3\%$ for $\nu=1.20$, $1.35$,
$1.45$ (table~\ref{tab:powerlaw_check}).

\begin{table}[t]
\caption{Numerical verification of the superhorizon power law
$|\mathcal{D}_k|\propto z^\nu$.  The ratio
$|\mathcal{D}_k(z_1)|/|\mathcal{D}_k(z_2)|$ at $z_1=0.003$,
$z_2=0.010$ is compared with $(z_1/z_2)^\nu$.}
\label{tab:powerlaw_check}
\begin{ruledtabular}\begin{tabular}{@{}cccc@{}}
$\nu$ & $m_\sigma/H$ & Numerical ratio & $(z_1/z_2)^\nu$ \\
\hline
1.45 & 0.55 & 0.17470 & 0.17451 \\
1.35 & 0.95 & 0.19694 & 0.19684 \\
1.20 & 1.32 & 0.23590 & 0.23580 \\
\end{tabular}\end{ruledtabular}
\end{table}

\begin{figure}[b]
\centering
\includegraphics[width=0.95\columnwidth]{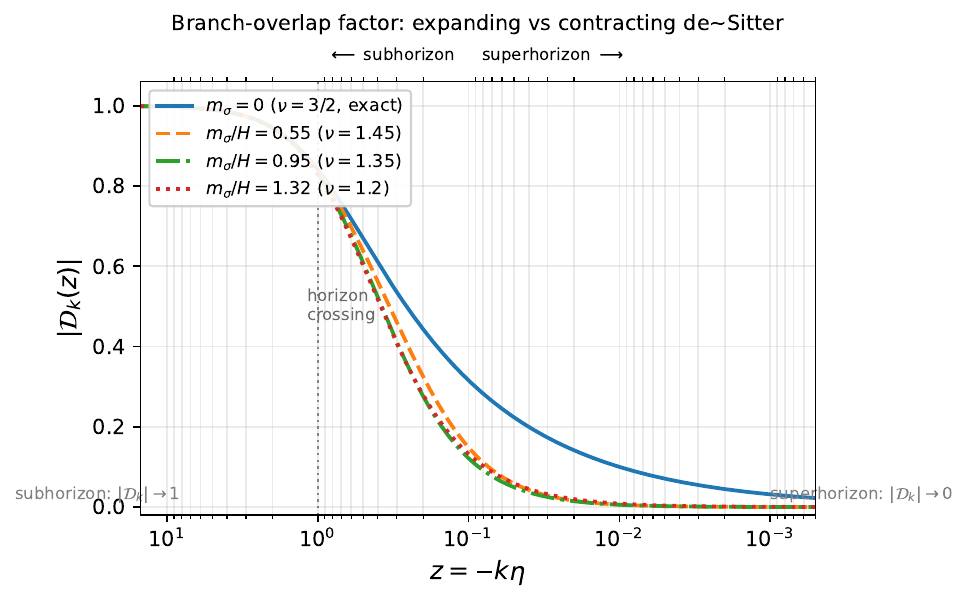}
\caption{Branch-overlap factor $|\mathcal{D}_k(z)|$ as a function of
$z=-k\eta$, from subhorizon ($z\gg1$) to superhorizon ($z\to0$), for
four spectator masses.  Curves are labelled by spectator mass
$m_\sigma/H$ (equivalently $\nu$); the massless case ($\nu=3/2$) uses
the exact closed form~(\ref{eq:D_massless_exact}); massive curves use
the general formula~(\ref{eq:D_overlap_result}) with Bessel functions
evaluated numerically.  All curves approach unity subhorizon and decay
to zero superhorizon.  The dotted vertical line marks horizon crossing
($z=1$).}
\label{fig:branch_overlap}
\end{figure}

The behavior of $|\mathcal{D}_k(z)|$ is illustrated in
figure~\ref{fig:branch_overlap}.

\subsubsection*{Accumulated branch-decoherence functional}

Converting the mode sum to a phase-space integral,
\begin{equation}
\Gamma_{+-}(\Nefold)
  = \frac{1}{2\pi^2}
    \int_0^{k_{\max}} \diff k\, k^2
    \left(-\ln|\mathcal{D}_k(z_k(\Nefold))|\right),
\label{eq:Gamma_integral}
\end{equation}
with $z_k(\Nefold)=k/(H\,\ee^{\Nefold})$.

\subsubsection*{Exact massless evaluation of the accumulated branch
functional}

While the single-mode overlap law is cutoff-independent, the accumulated
quantity $\Gamma_{+-}$ defined in equation~(\ref{eq:Gamma_integral})
necessarily depends on the chosen upper phase-space boundary
$k_{\max}$ once one sums over a finite mode band.  In the massless case
$\nu=3/2$, however, this dependence can be characterized exactly.
Using the closed-form overlap
\begin{equation}
|\mathcal{D}_k(z)|=\left[\frac{z^2}{z^2+1}\right]^{1/4},
\qquad
z_k(\Nefold)=\frac{k}{H\ee^{\Nefold}},
\label{eq:D_massless_exact_repeat}
\end{equation}
one has
\begin{equation}
-\ln |\mathcal{D}_k|
= \frac{1}{4}\ln\!\left(1+\frac{H^2\ee^{2\Nefold}}{k^2}\right).
\label{eq:minuslogD_massless}
\end{equation}
It is convenient to define the horizon scale
\begin{equation}
k_H(\Nefold)\equiv H\ee^{\Nefold}=a(\Nefold)H.
\label{eq:kH_def}
\end{equation}
Then equation~(\ref{eq:Gamma_integral}) becomes
\begin{equation}
\Gamma_{+-}^{(\nu=3/2)}(\Nefold;k_{\max})
=
\frac{1}{8\pi^2}
\int_0^{k_{\max}}\diff k\,k^2
\ln\!\left(1+\frac{k_H^2(\Nefold)}{k^2}\right).
\label{eq:Gamma_massless_integral}
\end{equation}
The primitive is elementary:
\begin{equation}
\begin{aligned}
\int \diff k\,k^2\ln\!\left(1+\frac{k_H^2}{k^2}\right)
&= \frac{k^3}{3}\ln\!\left(1+\frac{k_H^2}{k^2}\right)
  +\frac{2}{3}k_H^2 k
  \\
&\quad
  -\frac{2}{3}k_H^3\arctan\!\left(\frac{k}{k_H}\right),
\end{aligned}
\label{eq:Gamma_massless_primitive}
\end{equation}
and the lower endpoint $k\to0$ gives zero.  Therefore the accumulated
branch-decoherence functional in the massless case is
\begin{equation}
\begin{aligned}
\Gamma_{+-}^{(\nu=3/2)}(\Nefold;k_{\max})
&= \frac{1}{24\pi^2}\Biggl[
 k_{\max}^3\ln\!\left(1+\frac{k_H^2(\Nefold)}{k_{\max}^2}\right)
 \\
&\quad +2k_H^2(\Nefold)\,k_{\max}
 \\
&\quad -2k_H^3(\Nefold)
 \arctan\!\left(\frac{k_{\max}}{k_H(\Nefold)}\right)
\Biggr].
\end{aligned}
\label{eq:Gamma_massless_closed}
\end{equation}
Equation~(\ref{eq:Gamma_massless_closed}) replaces the purely numerical
integration in the exact de~Sitter massless case and makes explicit how
the accumulated branch separation is controlled by the ratio
$k_{\max}/k_H(\Nefold)$.

\subsubsection*{Asymptotic structure and cutoff dependence}

\begin{figure*}[t]
\centering
\includegraphics[width=\textwidth]{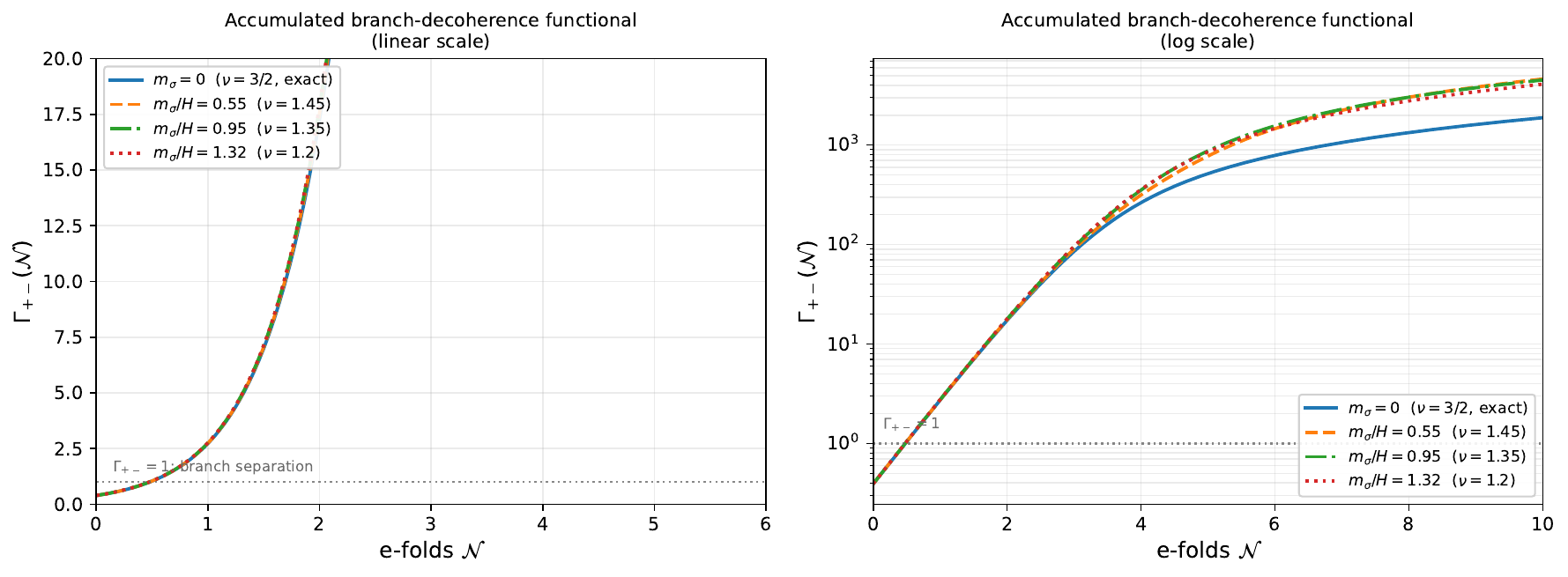}
\caption{Accumulated branch-decoherence functional $\Gamma_{+-}(\Nefold)$
on linear (left) and logarithmic (right) scales.  The dotted horizontal
line marks $\Gamma_{+-}=1$.  The threshold is crossed within $\approx0.5$
e-folds for all spectator masses; thereafter $\Gamma_{+-}$ grows
rapidly, confirming irreversible branch separation.  The massless curve
uses the exact closed form~(\ref{eq:D_massless_exact}); massive curves
use the general formula~(\ref{eq:D_overlap_result}).
This geometric $\Gamma_{+-}$ is independent of the coupling~$\hat\lambda$
and provides a lower bound on the decoherence timescale; the
coupling-dependent noise-kernel estimate yields a later crossing at
$\Nefold\approx7$ e-folds for $\hat\lambda=10^{-3}$
(table~\ref{tab:decoherence_times}).}
\label{fig:Gamma_efolds}
\end{figure*}

Equation~(\ref{eq:Gamma_massless_closed}) makes the ultraviolet structure
of the accumulated branch functional completely explicit.  For
$k_{\max}\gg k_H(\Nefold)$, expansion of the logarithm and arctangent gives
\begin{equation}
\begin{aligned}
\Gamma_{+-}^{(\nu=3/2)}(\Nefold;k_{\max})
&= \frac{k_H^2(\Nefold)\,k_{\max}}{8\pi^2}
 -\frac{k_H^3(\Nefold)}{24\pi}
 \\
&\quad
 +\frac{k_H^4(\Nefold)}{16\pi^2\,k_{\max}}
 +\mathcal{O}\!\left(\frac{k_H^6}{k_{\max}^3}\right).
\end{aligned}
\label{eq:Gamma_massless_asymptotic}
\end{equation}
In particular, the numerical crossing seen in
figure~\ref{fig:Gamma_efolds} should be interpreted as a property of the
chosen representative mode band, whereas the exact overlap law and the
coefficient of the universal subleading term are intrinsic.

The leading term is linear in the upper phase-space boundary and reflects
the accumulation of subhorizon mode contributions when the branch-overlap
suppression is summed over a finite mode band.  By contrast, the first
cutoff-independent term,
\begin{equation}
-\frac{k_H^3(\Nefold)}{24\pi},
\label{eq:Gamma_massless_universal_piece}
\end{equation}
is universal: its coefficient is fixed entirely by the exact single-mode
overlap law~(\ref{eq:D_massless_exact_repeat}) and is independent of the
choice of $k_{\max}$.  The exact massless calculation therefore separates
the accumulated branch functional into a cutoff-sensitive phase-space term
and a universal subleading contribution.  This analytic decomposition
sharpens the distinction between the coarse-grained EFT estimates of
section~\ref{subsec:coarsegraining}, whose rates depend explicitly on the
chosen environmental split, and the cutoff-independent single-mode
branch-overlap law derived in the present section.

Numerical evaluation using
equations~(\ref{eq:ReOmega})--(\ref{eq:D_massless_exact}) confirms
agreement with the closed form~(\ref{eq:Gamma_massless_closed}) to
machine precision.  For the general massive case, the integral must
still be evaluated numerically; the result shows that
$\Gamma_{+-}$ crosses unity within $\Nefold\approx0.5$ $e$-folds for
all values of $m_\sigma/H$ considered, and reaches $\mathcal{O}(10^3)$
within four to five $e$-folds (figure~\ref{fig:Gamma_efolds}).  Branch
separation is therefore not only efficient but effectively irreversible.

\medskip\noindent\textit{Two complementary decoherence rates.}---
It is important to distinguish the two accumulated functionals that
appear in this paper, which measure related but conceptually distinct
quantities and have different dependences on the UV cutoff.

\emph{(i)~Geometric branch-overlap functional $\Gamma_{+-}$.}
Figure~\ref{fig:Gamma_efolds} shows $\Gamma_{+-}(\Nefold)$ computed
directly from the geometric branch-overlap
formula~(\ref{eq:Gamma_integral}),
$\Gamma_{+-}=-\sum_k\ln|\mathcal{D}_k|$, using the Bunch--Davies mode
functions.  The \emph{single-mode} overlap factor $|\mathcal{D}_k(z)|$
is independent of both the spectator coupling~$\hat\lambda$ and the UV
cutoff.  The \emph{band-integrated} functional $\Gamma_{+-}$, however,
necessarily depends on the upper phase-space boundary~$k_{\max}$; in
the massless limit this dependence is analytically controlled by the
exact closed form~(\ref{eq:Gamma_massless_closed}) and the asymptotic
decomposition~(\ref{eq:Gamma_massless_asymptotic}), which separate
$\Gamma_{+-}$ into a cutoff-sensitive leading term and a universal
subleading piece.  The functional crosses unity within $\approx 0.5$
$e$-folds for all spectator masses considered in a representative
mode band.

\emph{(ii)~Noise-kernel functional
(cutoff-dependent).}  The coupling-dependent noise-kernel estimate of
section~\ref{subsec:coarsegraining}, by contrast, depends explicitly on
$\hat\lambda$, on the branch-separation amplitude
$|\Delta\zeta_{k_L}|$, and on the UV regulation of the noise
kernel~$N(x,x')$.  The loop integral defining~$N$ requires a UV cutoff
or dimensional-regularization scheme, and the resulting decoherence
timescale inherits that dependence at the level of the numerical
prefactor.  For $\hat\lambda=10^{-3}$ the scalar threshold is crossed
at $\Nefold\approx 7$ $e$-folds
(table~\ref{tab:decoherence_times}).

The two calculations are complementary.
The geometric $\Gamma_{+-}$ is built from the single-mode overlap law,
which is a purely kinematic consequence of inflationary squeezing and is
independent of any coupling or cutoff; the band-integrated diagnostic
inherits a $k_{\max}$-dependence whose analytic structure is fully
characterized by
equations~(\ref{eq:Gamma_massless_closed})--(\ref{eq:Gamma_massless_asymptotic}).
The noise-kernel estimate quantifies the rate at which a specific
spectator sector exploits the branch asymmetry to suppress off-diagonal
density-matrix elements, which requires the sustained accumulation of
environmental records over several $e$-folds and inherits the UV
sensitivity of the noise kernel.  Both calculations confirm irreversible
branch separation well within inflation.

\subsubsection*{Physical interpretation}

Equation~(\ref{eq:D_overlap_result}) makes the mechanism transparent.
The overlap is governed by the squeezing angle $\theta_k=\arg\Omega_k^{(+)}$
of the expanding-branch vacuum.  With $R>0$ and $I<0$ (equation~(\ref{eq:ReOmega})),
one has $\tan\theta_k = I/R = -1/z$, so $\theta_k\in(-\pi/2,0)$ for all
$z>0$.  Subhorizon modes have $z\gg1$, giving $|\theta_k|\approx0$ and
$|\mathcal{D}_k|\approx1$ --- the two branches are indistinguishable.
As a mode crosses the horizon, inflationary particle production drives
$z\to0$, so $|\theta_k|\to\pi/2$ and $R=kz^2/(z^2+1)\to0$, giving
$|\mathcal{D}_k|\to0$.  On the contracting branch the mode function
$u_k^{(-)}$ does not undergo the same superhorizon amplification, so
the squeezing angle on that branch remains $\mathcal{O}(1)$; the
two-branch overlap therefore vanishes because of the asymmetry
between $\Omega_k^{(+)}$ and $\Omega_k^{(-)}$.  The asymmetry is
a direct consequence of the asymmetry in superhorizon mode production
--- the same physical process that generates the observed scale-invariant
curvature spectrum.  Branch decoherence and the generation of classical
primordial perturbations are two facets of the same inflationary
squeezing.  The same asymmetry between expanding and contracting
branches applies in bouncing cosmology scenarios where a
Wheeler--DeWitt contracting phase precedes
inflation~\cite{Demetrio2026}.

\paragraph{Corollary: universality for tensor mode functions in exact de~Sitter.}
In exact de~Sitter space, each polarization of the transverse-traceless
tensor perturbation obeys the same mode equation as a massless minimally
coupled scalar, with index $\nu=3/2$.  At the level of the geometric
branch-overlap calculation considered here, the tensor mode functions
therefore lead to the same single-mode overlap law as the massless scalar,
namely
\begin{equation}
|\mathcal{D}^{(T)}_k(z)|
=
\left[\frac{z^2}{z^2+1}\right]^{1/4},
\qquad
z=-k\eta .
\label{eq:D_tensor_corollary}
\end{equation}
Thus the exact de~Sitter branch-overlap factor is universal across
massless scalar and tensor mode functions.  If one defines an accumulated
geometric branch functional by summing over tensor modes in the same
representative comoving band as in equation~(\ref{eq:Gamma_integral}),
the result differs only by the trivial multiplicity factor of two coming
from the two tensor polarizations.  We emphasize, however, that this
statement concerns only the cutoff-independent geometric overlap law.
A full comparison of scalar and tensor \emph{decoherence rates} would
require a separate analysis of the relevant system--environment couplings
and is not pursued here.

\subsection{Relation to previous work}

There are two related but distinct literatures to compare with the
present calculation.  The first concerns decoherence between
semiclassical branches in quantum cosmology.  Kiefer's early analysis
already emphasized that WKB components corresponding to expanding and
contracting universes can decohere through their coupling to
perturbative degrees of freedom~\cite{Kiefer1992}.  Later work by
Kiefer and collaborators developed the reduced-density-matrix viewpoint
and its connection with the emergence of classical spacetime and
classical perturbations~\cite{Kiefer,KieferPolarski,KieferBook}.
Conceptually, the present paper follows this line: classicality is a
property of a reduced state, not of the WKB approximation alone.
Technically, the difference is that we compute the branch overlap
directly from the inflationary Gaussian kernels and Bunch--Davies mode
functions, yielding the closed single-mode expression
(\ref{eq:D_overlap_result}) and the closed massless accumulated
functional~(\ref{eq:Gamma_massless_closed}).  These formulae make the
expanding--contracting suppression quantitative in a form that can be
compared mode by mode with the usual squeezing variables.

The second literature concerns decoherence and classicality of
primordial perturbations during inflation.  Polarski and Starobinsky
emphasized the semiclassical appearance of squeezed cosmological
perturbations and the role of decoherence in rendering them effectively
classical~\cite{PolarskiStarobinsky}.  Burgess, Holman, and Hoover
analyzed the decoherence of inflationary fluctuations from the
perspective of effective field theory and subhorizon environmental
interactions~\cite{BurgessHolman}, with subsequent work extending the
analysis to the open-EFT framework~\cite{BurgessHolmanStamou2022}.  The
quantum discord and einselection approach to inflationary classicality
has been developed in~\cite{Zurek2009,MartinVenninPeter2012}.

Recent work has also stressed that squeezing and environmental
decoherence do not necessarily erase every possible quantum signature of
primordial fluctuations.  In particular, Martin and Vennin argued that
some inflationary quantum correlations can remain, in principle,
operationally meaningful even when the perturbations look classical in
standard cosmological observables~\cite{MartinVennin2017}.  This is not
in conflict with the present result.  Our claim is the narrower
open-system claim that off-diagonal branch matrix elements are strongly
suppressed in a specified coarse graining.  It does not require that all
quantum discord, Bell-type nonclassicality, or fine-grained information
in the full universal state vanish.  The exact state remains pure under
unitary evolution; decoherence explains the practical diagonality and
record stability of the branch-relative reduced state.

The main point of the present analysis is therefore not to propose a
wholly new decoherence mechanism, but to assemble a more explicit bridge
between ingredients that are often discussed separately: the origin of
the effective long-mode coupling to spectator sectors, the
coarse-graining dependence of the resulting decoherence exponent, the
closed-form branch-overlap diagnostic, and the different roles played by
Hartle--Hawking and tunneling boundary conditions.

\subsection{Boundary conditions versus classicality}

The Hartle--Hawking and tunneling proposals determine the relative
amplitudes~$A_{\alpha}$ of semiclassical branches, but they do so in
different ways: the no-boundary state gives a real standing-wave
combination, while the tunneling prescription gives an outgoing complex
WKB branch.  Decoherence determines when the overlap between branch
conditioned environmental states becomes negligibly small.  Without
decoherence, expanding and contracting branches remain part of a
coherent superposition in the Hartle--Hawking state; with decoherence,
the branch-overlap factor~$\mathcal{D}(a)$
in~(\ref{eq:branch_overlap_general}) determines when the two histories
can be treated as an effectively classical ensemble.  For the tunneling
state, by contrast, the outgoing WKB form already encodes the branch
selection; decoherence then explains the robustness of the selected
semiclassical history rather than selecting it from an equal real
superposition.  The explicit calculation in section~\ref{sec:boundary}
establishes that for the inflationary coarse graining adopted here, the
overlap $|\mathcal{D}_k|$ between the expanding- and
contracting-branch conditioned states tends to zero as modes freeze out
superhorizon, driven by the asymmetry in superhorizon squeezing between
the two branches.  This is a derived result, not an
inference: it follows directly from the squeezing asymmetry encoded in
equations~(\ref{eq:D_overlap_result})
and~(\ref{eq:D_massless_exact}).

\subsection{Decoherence and the arrow of time}

The arrow-of-time claim made here is deliberately limited.  Neither the
Wheeler--DeWitt equation nor the reduced density matrix of the exact
closed universe contains a fundamental external time parameter or a
microscopic violation of time-reversal invariance.  The problem is
instead to explain why observers inside one semiclassical branch see
stable records, retarded stochastic evolution, and a monotonic growth of
accessible entropy in one temporal orientation.  In the terminology of
Zeh, decoherence does not create a fundamental arrow from nothing; it
amplifies the time asymmetry implicit in the special cosmological
boundary condition into an operational arrow carried by records and
correlations~\cite{ZehBook}.  This perspective is complementary to
Penrose's low-entropy initial-condition viewpoint, in particular the
Weyl-curvature hypothesis, which identifies the special initial
geometric state as the origin of the thermodynamic arrow~\cite{Penrose1979}.

In the present calculation that amplification is explicit.  Inflation
provides a sequence of mode crossings.  For each mode, squeezing drives
the phase-space ellipse toward the superhorizon limit, and the
conditioned environmental states associated with opposite WKB
orientations become less and less overlapping.  The exact single-mode
law~(\ref{eq:D_overlap_result}) gives
$|\mathcal{D}_k|\rightarrow0$ as $z=-k\eta\rightarrow0$ in the massless
case, while the massive result gives the corresponding power-law
suppression.  Summing over the modes in a coarse-grained band gives the
monotonically increasing $\Gamma_{+-}(\Nefold)$ shown in
figure~\ref{fig:Gamma_efolds}.  The branch overlap becomes negligible
within the first $e$-fold for the representative bands considered
there, and subsequent horizon crossings add further records rather than
restoring coherence.

The resulting arrow is therefore a branch-relative, coarse-grained
arrow.  In the Hartle--Hawking case, decoherence converts a real
superposition of expanding and contracting WKB components into an
effectively classical ensemble whose branches no longer interfere.  In
the tunneling case, the outgoing WKB boundary condition already selects
the expanding orientation, and decoherence explains why that orientation
supports persistent classical records.  In both cases, the mechanism is
not the boundary amplitude alone and not squeezing alone, but the
combination of a special quantum-cosmological state, inflationary
squeezing, and environmental coarse graining.

\section{Discussion and outlook}
\label{sec:discussion}

We have demonstrated that environment-induced decoherence provides an
efficient and robust mechanism for the dynamical emergence of classical
cosmology.  The influence-functional framework is consistent with the
stochastic description of inflationary
perturbations~\cite{StarobinskyStochastic,NambuSasaki1989}, and our
results clarify the conceptual foundations of quantum cosmology by
providing a concrete bridge between universal wavefunctions and
classical spacetime histories.

The exact analytic evaluation of $\Gamma_{+-}$ in the massless limit
(equation~(\ref{eq:Gamma_massless_closed})) and its asymptotic
decomposition (equation~(\ref{eq:Gamma_massless_asymptotic})) provide complete
analytic control over the accumulated branch functional, including an
explicit separation into cutoff-dependent and universal pieces.  The
extension to tensor perturbations
(equation~(\ref{eq:D_tensor_corollary})) confirms the universality of the
overlap law for all massless de~Sitter modes.

Future directions include extending the analysis to the reheating
epoch, computing quantum-information-theoretic measures of
cosmological decoherence --- in particular the relative entropy or
Bures fidelity between the two branch-conditioned Gaussian states,
which are directly computable from the overlap
factor~(\ref{eq:D_overlap_result}) --- and connecting the tensor
overlap-law universality established in the corollary of
section~\ref{sec:boundary} to primordial $B$-mode observables via the
tensor-to-scalar ratio~$r$, which would require a separate treatment of
the relevant system--environment couplings for tensor perturbations.

\section{Conclusion}
\label{sec:conclusion}

In this work we have studied how classical spacetime and classical
cosmological perturbations emerge from quantum cosmology.  Starting
from standard quantum-cosmological boundary conditions, including the
no-boundary and tunneling proposals, we emphasized that semiclassical
WKB behavior and squeezing of perturbations are not by themselves
sufficient to explain classicality.

By explicitly tracing over unobserved degrees of freedom and employing
the influence functional formalism, we derived a decoherence functional
for long-wavelength curvature perturbations during inflation.  We showed
that interactions with environmental degrees of freedom lead to an
efficient and robust suppression of quantum interference between
macroscopically distinct perturbation histories.  This mechanism renders
the reduced density matrix approximately diagonal and justifies the use
of classical stochastic descriptions in cosmology.

We demonstrated that the emergence of classicality is largely
insensitive to the details of coarse graining, occurring both for
horizon-based and EFT-motivated splits between system and environment.
Classical cosmological histories therefore arise as a generic
consequence of inflationary dynamics in the presence of additional
degrees of freedom, rather than as a fine-tuned outcome of specific
assumptions.

Our analysis clarifies the distinct roles played by quantum-cosmological
boundary conditions and decoherence.  Boundary conditions determine the
relative amplitudes of semiclassical branches in the universal
wavefunction, while decoherence determines when the overlap between
these branches becomes negligibly small.  The explicit evaluation of the
branch-overlap factor $|\mathcal{D}_k(z)|$ shows that the overlap
between expanding- and contracting-branch conditioned states
vanishes in the superhorizon limit --- as a
power law $z^\nu$ for massive fields and as $z^{1/2}$ exactly for the
massless case --- driven by the asymmetry in superhorizon squeezing.
In the massless limit the accumulated functional $\Gamma_{+-}$ can be
evaluated in exact closed form, and its asymptotic expansion separates
cleanly into a cutoff-sensitive leading term and a universal subleading
piece.  The functional crosses the classicality threshold within half an
$e$-fold for representative mode bands and grows without bound.  This
supports a branch-relative arrow of time: inflationary squeezing and
environmental coarse graining create stable records along one
semiclassical orientation without requiring a fundamental arrow to be
inserted into the Wheeler--DeWitt equation.

Taken together, these results provide a more explicit framework
connecting quantum cosmology, decoherence, and the classical universe
described by cosmological observations.  In particular, they sharpen
how coarse graining and branch decoherence enter the passage from a
universal wavefunction to an effective ensemble of classical
cosmological histories.

\begin{acknowledgments}
The author thanks Luiz Felipe Dem\'etrio for correspondence regarding
bouncing cosmology and Wheeler--DeWitt quantum bounce scenarios, which
motivated the discussion of branch decoherence in the context of
contracting-phase cosmologies in section~\ref{sec:boundary}.
\end{acknowledgments}

\appendix
\section{Derivation of the decoherence time estimate}
\label{app:derivation}

For convenience we record the intermediate steps leading from the
EFT-based coarse-grained estimate to the explicit decoherence times in
table~\ref{tab:decoherence_times}.  Starting from
\begin{equation}
\frac{\diff\Gamma_{k_L}}{\diff\Nefold}
  \sim \frac{\lambda_{\mathrm{eff}}^{2}\Lambda_{\mathrm{phys}}}{H^{5}}\,
       |\Delta\zeta_{k_L}|^2\,a^4(\Nefold),
\end{equation}
and writing $a(\Nefold)=a_\ast\,\ee^{\Delta\Nefold}$ relative to
horizon exit of the long mode, one finds
\begin{equation}
\frac{\diff\Gamma_{k_L}}{\diff\Nefold}
  \sim \frac{\lambda_{\mathrm{eff}}^{2}\Lambda_{\mathrm{phys}}}{H^{5}}\,
       |\Delta\zeta_{k_L}|^2\,a_\ast^4\,\ee^{4\Delta\Nefold}.
\end{equation}
Using horizon exit, $a_\ast H=k_L$, as the reference scale and
absorbing order-unity factors into the normalization,
\begin{equation}
\frac{\diff\Gamma_{k_L}}{\diff\Nefold}
  \sim \hat\lambda^2\,\frac{\Lambda_{\mathrm{phys}}}{H}\,
       |\Delta\zeta_{k_L}|^2\,\ee^{4\Delta\Nefold},
\quad
\hat\lambda\equiv\frac{\lambda_{\mathrm{eff}}}{H^2}.
\end{equation}
Integrating from $\Delta\Nefold'=0$ to $\Delta\Nefold$ yields
\begin{equation}
\Gamma_{k_L}(\Delta\Nefold)
  = \frac{\hat\lambda^2}{4}\,\frac{\Lambda_{\mathrm{phys}}}{H}\,
    |\Delta\zeta_{k_L}|^2\,
    \left(\ee^{4\Delta\Nefold}-1\right),
\end{equation}
which is the expression used in
equation~(\ref{eq:decoherence_efolds_exact}).  Solving for
$\Delta\Nefold_{\mathrm{dec}}$ subject to
$\Gamma_{k_L}(\Delta\Nefold_{\mathrm{dec}})\simeq1$ reproduces
equation~(\ref{eq:decoherence_efolds_exact}).  Using
$\zeta_{\mathrm{rms}}\simeq\sqrt{A_s}\simeq5\times10^{-5}$ and
$\hat\lambda=\tfrac{3}{2}m_\sigma^2/H^2$ yields the numerical values
in table~\ref{tab:decoherence_times} and the mass-based examples in
section~\ref{subsec:estimates}.


\end{document}